Technical Notes

# The Omitted Noise Contribution of Surface Normal Variation: Farassat's Formulation 1A revisited

Qitian Tao[1], Chen He[*2], Xiaohua Liu[*1,2], Zhengwu Chen[2], Jiale Lu[1]

1. School of Aeronautics and Astronautics, Shanghai Jiao Tong University, Shanghai, 200240, China
2. State Key Laboratory of Aerodynamics, China Aerodynamics Research and Development Center, Mianyang, Sichuan, 621000, China

* Corresponding author, E-mails: yushanhechen@163.com; Xiaohua-Liu@sjtu.edu.cn

## 1 Introduction

Lighthill's acoustic analogy[1] describes sound radiation from compact turbulent regions in an unbounded, homogeneous medium. The continuity and momentum equations are reformulated into an inhomogeneous wave equation, which establishes an exact far-field analogy between the density fluctuations in a real flow and the small amplitude density fluctuations generated by an equivalent quadrupole source distribution in a fictitious stationary acoustic medium. It is often possible to obtain reasonably accurate estimates of the quadrupole source in certain types of flows, and consequently, correspondingly accurate predictions of the resulting sound field.

The acoustic analogy has played an important role in the prediction of rotating machinery noise. In such problems, solid boundaries support surface distributions of monopole and dipole noise sources, which arise from the motions and forces imparted to the surface by the unsteady flow. The Ffowcs Williams-Hawkings (FW-H) formulation[2] extends Lighthill's acoustic analogy by incorporating the effects of moving surfaces, leading to a generalized representation of the acoustic field through the outgoing Green's function.

By decoupling the aerodynamic and acoustic fields, the acoustic analogy greatly simplifies aeroacoustic noise prediction and enables the use of linear analysis tools to solve the resulting linear acoustic problem. The FW-H equation treats the acoustic field as a superposition of elementary wave solutions. Volume elements outside the control surface radiate sound as quadrupole sources, in agreement with Lighthill's acoustic analogy. Surface elements on the control surface generate equivalent dipole radiation associated with the unsteady loading acting on the solid boundary. Volume elements within the control surface produce a combined dipole-quadrupole contribution that is mathematically equivalent to a monopole source, arising from the volume displacement effect associated with boundary motion.

The prediction of the acoustic pressure field using the FW-H equation requires accurate representations of the surface geometry and motion, the gauge pressure on the moving surface, and the Lighthill stress tensor in the surrounding exterior volume. Farassat developed several time-domain solutions of the FW-H equation that have significantly contributed to propeller noise prediction. By neglecting the quadrupole source terms, Formulation 1[3] was first introduced to address noise generation from helicopter rotors, and a much simpler derivation was later presented[4]. Formulation 1A[5] is derived using the Green's function of the wave equation, with reduced computational cost and improved accuracy. Farassat[6] subsequently summarized his work, and these formulations are widely regarded as some of the most effective approaches for solving the FW-H equation.

The loading noise represents the acoustic radiation generated by unsteady forces associated with the interaction between the source surface and the surrounding fluid. When viscous stress contributions are neglected, the unsteady force acts normal to the source surface, with a magnitude determined by the surface gauge pressure. However, in the derivation from Formulation 1 to Formulation 1A in Refs. [5,6], the contribution associated with the temporal variation of the direction of the unsteady force is omitted, which mathematically manifests itself as the temporal derivative of the local surface normal vector (TDNV). Although this term does not affect the far-field acoustic prediction of stationary or uniformly translating sources, its omission may introduce errors in rotating machinery noise predictions. This mathematical oversight motivates a revisit of the Formulation 1A, and the contribution of surface normal variation to far-field loading noise is therefore re-examined in the present study.

In this study, Formulation 1A is rigorously rederived to demonstrate the existence of the TDNV contribution, and the physical significance of each term is clarified. The resulting formulation, referred to as the Modified Formulation 1A, is compared with the original Formulation 1A. The physical applicability of the two formulations is discussed, and the validity of the Modified Formulation 1A proposed in the present study is demonstrated through a propeller far-field noise case study.

## 2 On the Derivation and Impact of the TDNV Term

As a generalization of Lighthill's acoustic analogy for moving solid boundaries, the FW-H equation[2] is an inhomogeneous wave equation, given by

$$\Box^2 p' = \frac{\partial}{\partial \tau}\left[\rho_0 v_n \delta(f)\right] - \frac{\partial}{\partial x_i}\left[p n_i \delta(f)\right] + \frac{\partial^2}{\partial x_i \partial x_j}\left[H(f) T_{ij}\right] \tag{1}$$

where $p'$ is the aeroacoustic pressure, $\tau$ is the source time, $\rho_0$ is the ambient density, $v_n$ is the normal velocity of the source surface, $p$ is the surface gauge pressure, and $n_i$ are the components of the unit normal vector. $T_{ij}$ denotes the Lighthill stress tensor, and $\Box^2$ denotes the three-dimensional wave operator. $H(f)$ and $\delta(f)$ represent the Heaviside and Dirac delta functions, respectively.

The differential operator in the FW-H equation is linear with respect to the acoustic pressure perturbation, whereas the source terms are fully determined by flow variables independent of the acoustic field. It therefore constitutes a linear wave equation that satisfies the superposition principle. Consequently, the acoustic contributions from different source mechanisms can be evaluated independently and superposed linearly. The loading noise component $p_L'$ can be written as

$$\Box^2 p_L' = -\frac{\partial}{\partial x_i}\left[p n_i \delta(f)\right] \tag{2}$$

Through a transformation of the FW-H equation, Farassat derived a new integral formulation for subsonically moving surfaces, referred to as Formulation 1[7]. The loading noise component of Formulation 1 is expressed as

$$4\pi p_L' = \frac{1}{c}\frac{\partial}{\partial t}\int_{f=0}\left[\frac{p\cos\theta}{r\left|1-M_r\right|}\right]_{ret} dS + \int_{f=0}\left[\frac{p\cos\theta}{r^2\left|1-M_r\right|}\right]_{ret} dS \tag{3}$$

where $t$ is the observer time, $\theta$ denotes the instantaneous angle between the surface normal vector and the radiation direction, and $\cos\theta = n_i \hat{r}_i$. Here, $\hat{r}_i$ are the components of the unit radiation vector, $M_r$ is the Mach number in the radiation direction, and $r$ is the source-observer distance. The subscript 'ret' stands for retarded time, and $c$ is the speed of sound in the undisturbed medium.

Formulation 1 contains an observer time derivative that must be evaluated numerically, potentially introducing additional numerical errors. Formulation 1A[7] eliminates this issue by transforming it into a source time derivative, yielding

$$4\pi p_L' = \int_{f=0} \left[ \frac{\dot{p}\cos\theta}{cr(1-M_r)^2} + \frac{p\cos\theta \hat{r}_i \dot{M}_i}{cr(1-M_r)^3} + \frac{p\cos\theta(M_r - M^2)}{r^2(1-M_r)^3} + \frac{p(\cos\theta - M_i n_i)}{r^2(1-M_r)^2} \right]_{\mathrm{ret}} dS \tag{4}$$

where $\dot{q}$ denotes differentiation of a quantity $q$ with respect to source time $\tau$, $M$ is the Mach number of the source surface, and $M_i$ denote the components of the Mach number.

The contribution of loading noise is governed by the temporal variations in both the magnitude and direction of the unsteady surface force. Although each term in Formulation 1A admits a clear physical interpretation, the temporal variation in surface force direction is not explicitly represented. As discussed previously, when viscous contributions are neglected, the surface force acts in the direction of the local surface normal vector. Consequently, the governing equation should, in principle, include the temporal variation of the surface normal vector $\dot{\boldsymbol{n}}$. However, this contribution is absent from Formulation 1A.

Using Formulation 1 (Eq. 3), the observer time derivative is moved inside the first integral following a procedure analogous to that proposed by Farassat, yielding

$$4\pi p_L' = \int_{f=0} \left\{ \frac{1}{1-M_r} \frac{\partial}{\partial\tau} \left[ \frac{p\cos\theta}{cr|1-M_r|} \right] \right\}_{ret} dS + \int_{f=0} \left[ \frac{p\cos\theta}{r^2|1-M_r|} \right]_{ret} dS \tag{5}$$

The term involving the temporal derivative in Eq. (5) can be written as

$$\frac{1}{1-M_r} \frac{\partial}{\partial\tau} \left[ \frac{p\cos\theta}{cr(1-M_r)} \right] = \frac{\frac{\partial}{\partial\tau}(p\cos\theta)}{cr(1-M_r)^2} + \frac{p\cos\theta}{c(1-M_r)} \left[ \frac{\partial}{\partial\tau} \frac{1}{r(1-M_r)} \right] \tag{6}$$

It is worth noting that $\cos\theta = n_i \hat{r}_i$, as discussed above. If $n_i$ is treated as a constant during the derivation, the result of Formulation 1A can be fully recovered. However, for a generally moving source surface, the local surface normal vector is inherently time-dependent, as illustrated in Fig. 1. Without loss of generality, the temporal derivative of $p\cos\theta$ can be written as

$$\frac{\partial}{\partial\tau}(p\cos\theta) = \dot{p}\cos\theta + p\frac{\partial}{\partial\tau}(\cos\theta)$$
$$= \dot{p}\cos\theta + p\dot{n}_i\hat{r}_i + pn_i\dot{\hat{r}}_i \quad (7)$$

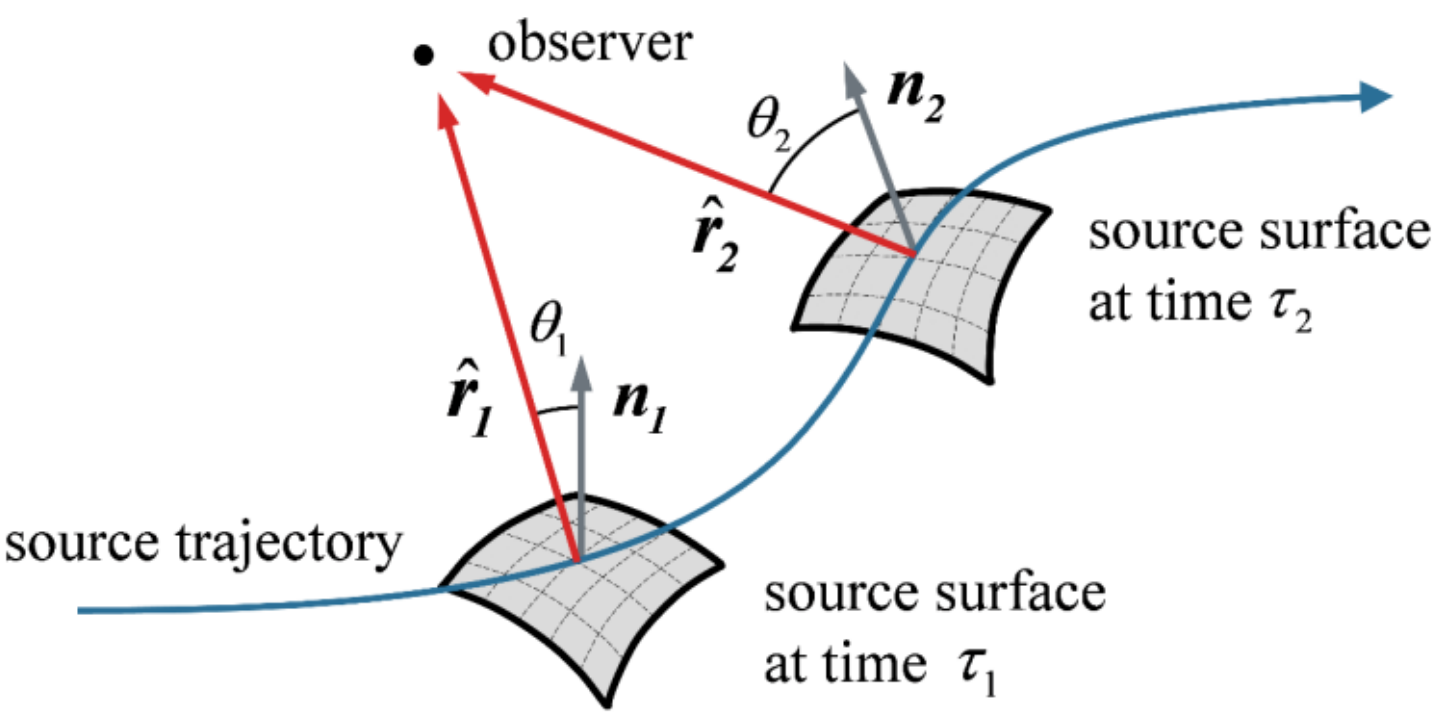

Fig. 1 Schematic of the source surface trajectory

For moving surfaces, vector algebra yields the following relation

$$\dot{\hat{r}}_i = \frac{\partial\hat{r}_i}{\partial\tau} = \frac{\hat{r}_i v_r - v_i}{r} \quad (8)$$

where $v_i$ are the components of the velocity vector, and $v_r$ is the source velocity in the radiation direction.

Equation (7) can then be further expanded as

$$\frac{\partial}{\partial\tau}(p\cos\theta) = \dot{p}\cos\theta + p\dot{n}_i\hat{r}_i + \frac{pn_i\left(\hat{r}_i v_r - v_i\right)}{r} \quad (9)$$

Substituting Eq. (9) into Eq. (6), the temporal derivative term can be rewritten as

$$\frac{1}{1-M_r}\frac{\partial}{\partial\tau}\left[\frac{p\cos\theta}{cr\left(1-M_r\right)}\right] = \frac{\dot{p}\cos\theta}{cr\left(1-M_r\right)^2} + \frac{p\dot{n}_i\hat{r}_i}{cr\left(1-M_r\right)^2} + \frac{pn_i\left(\hat{r}_i v_r - v_i\right)}{cr^2\left(1-M_r\right)^2}$$
$$+ \frac{p\cos\theta\hat{r}_i\dot{M}_i}{cr\left(1-M_r\right)^3} + \frac{p\cos\theta\left(M_r - M^2\right)}{r^2\left(1-M_r\right)^3} \quad (10)$$

Therefore, combining Eqs. (5) and (10), the loading noise component can be rewritten as

$$4\pi p_L' = \int_{f=0}\left[\frac{\dot{p}\cos\theta}{cr\left(1-M_r\right)^2} + \frac{p\cos\theta\hat{r}_i\dot{M}_i}{cr\left(1-M_r\right)^3} + \frac{p\cos\theta\left(M_r - M^2\right)}{r^2\left(1-M_r\right)^3}\right.$$
$$\left. + \frac{p\left(\cos\theta - M_i n_i\right)}{r^2\left(1-M_r\right)^2} + \frac{p\dot{n}_i\hat{r}_i}{cr\left(1-M_r\right)^2}\right]_{ret} dS \quad (11)$$

Equation (11), referred to herein as the Modified Formulation 1A, is derived from Formulation 1 (Eq. 3) and represents a time-domain integral solution of the FW-H equation for moving surfaces. It is theoretically equivalent to Formulation 1 in terms of predicted results. Compared with original Formulation 1A (Eq. 4), an additional contribution associated with the temporal derivative of the local surface normal vector is retained $\frac{p\dot{n}_i\hat{r}_i}{cr\left(1-M_r\right)^2}$. This term reflects the temporal variation in the direction of the surface force. Importantly, the existence of this term follows rigorously from FW-H equation.

Each term in the Modified Formulation 1A admits a clear physical interpretation associated with a distinct physical mechanism: Term 1 corresponds to the temporal derivative of the static pressure on moving surfaces (TDP); Terms 2 and 3 are associated with the temporal derivative of the surface Mach number (TDMA); Term 4 accounts for the temporal derivative of the radiation direction (TDR); and Term 5 corresponds to the temporal derivative of the surface normal vector (TDNV).

Alternatively, based on the asymptotic order with respect to the source-observer distance $r$, Modified Formulation 1A can be decomposed into far-field and near-field terms. Terms 1, 2, and 5 are of order $r^{-1}$ and therefore constitute the far-field terms, whereas Terms 3 and 4 are of order $r^{-2}$ and correspond to the near-field terms. This implies that the omission of Term 5 (TDNV) has a more significant impact on far-field loading noise prediction than Terms 3 and 4.

## 3 Validation and Discussion

The magnitude of the TDNV depends on the kinematic characteristics of the source surface. For non-rigid surfaces, structural deformation and vibration may render the unit normal vector $\boldsymbol{n}$ inherently time-dependent, making its contribution potentially significant in far-field noise prediction.

For rigid surfaces, without loss of generality, consider a non-inertial reference frame attached to the source surface, translating with velocity $\boldsymbol{v}_0(\tau)$ and rotating with angular velocity $\boldsymbol{\omega}(\tau)$. The velocity $\boldsymbol{V}(\boldsymbol{\xi},\tau)$ of an arbitrary point $\boldsymbol{\xi}$ in this frame can be expressed as

$$\boldsymbol{V}(\boldsymbol{\xi},\tau)=\boldsymbol{v}_0(\tau)+\boldsymbol{\omega}(\tau)\times\boldsymbol{\xi} \tag{12}$$

In this case, the temporal variation of the surface normal vector arises solely from rotational motion. Only when $\boldsymbol{\omega}(\tau)=0$, i.e., in the absence of rotation, does the normal vector remain time-invariant, and the TDNV contribution vanishes. Under this condition, the present Modified Formulation 1A reduces exactly to the original Formulation 1A.

For stationary or purely translating rigid-boundary noise problems, such as flow past a circular cylinder, sound generation is dominated by flow separation and periodic vortex shedding. Since the body geometry remains unchanged and no surface-normal variation is involved, the TDNV contribution vanishes, and accurate noise predictions can therefore be achieved using Formulation 1A. In contrast, for rotating rigid-boundary noise problems, such as the aerodynamic noise generated by a rotating propeller, the loading source is governed by unsteady aerodynamic forces acting on the moving blade surface. In such cases, the relative motion between the source surface and the observer renders the TDNV contribution non-negligible. Consequently, inclusion of the TDNV term extends the applicability of Formulation 1A from stationary or purely translating rigid surfaces to arbitrarily moving bodies undergoing general three-dimensional motion.

Meanwhile, a theoretical analysis of the magnitude of the TDNV term is performed as follows. The ratio of the TDNV term to Term 2 in the Modified Formulation 1A (Eq. 11) can be approximated by the ratio between the temporal derivative of the surface normal vector projected onto the radiation direction and that of the Mach number projected onto the same direction, yielding

$$\frac{p\dot{n}_i\hat{r}_i}{cr\left(1-M_r\right)^2} \Bigg/ \frac{p\cos\theta\hat{r}_i\dot{M}_i}{cr\left(1-M_r\right)^3} = \frac{\left(1-M_r\right)\dot{n}_i\hat{r}_i}{\cos\theta\dot{M}_i\hat{r}_i} \approx K\frac{\dot{n}_i\hat{r}_i}{\dot{M}_i\hat{r}_i} \tag{13}$$

where the coefficient $K$ can be regarded as a nonzero constant associated with the source-motion characteristics. Since this ratio is generally non-negligible, the contribution of the TDNV term to the far-field acoustic pressure is expected to be comparable to that of the other source terms. The TDNV term should therefore be retained in both theoretical formulations and numerical implementations.

Furthermore, a numerical validation of the TDNV contribution is conducted. In the present study, a computational solver is developed for predicting the far-field loading noise generated by rotating propeller blades. The acoustic pressure signals are evaluated using Farassat's Formulation 1, Formulation 1A, and the present Modified Formulation 1A, respectively. As illustrated in Fig. 2, an 11-bladed single-rotation propeller is considered as the test case. The blade surface pressure distributions, serving as the steady loading noise sources, are obtained from Reynolds-Averaged Navier–Stokes (RANS) simulations.

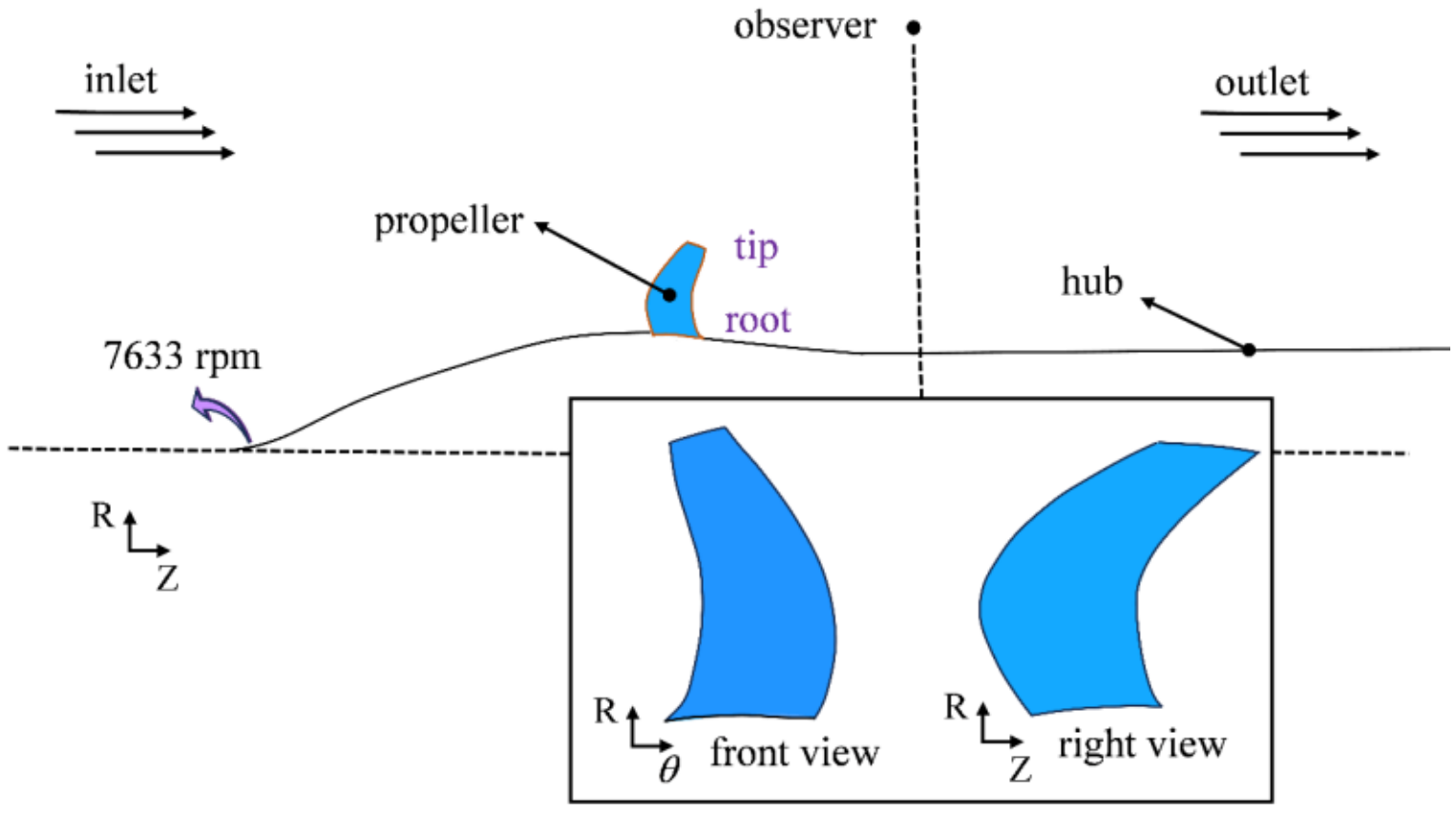

Fig. 2 Configuration of the propeller

With identical input parameters applied to all formulations, the predicted acoustic pressure signals are presented in Fig. 3. For the steady loading noise component, Formulation 1A exhibits a noticeable discrepancy relative to the present Modified Formulation 1A because the absence of the temporal derivative term associated with the source surface normal vector. In the present propeller case, the TDNV contribution accounts for approximately 41.5% of the loading noise amplitude. Meanwhile, compared with Formulation 1, the present Modified Formulation 1A achieves satisfactory prediction accuracy.

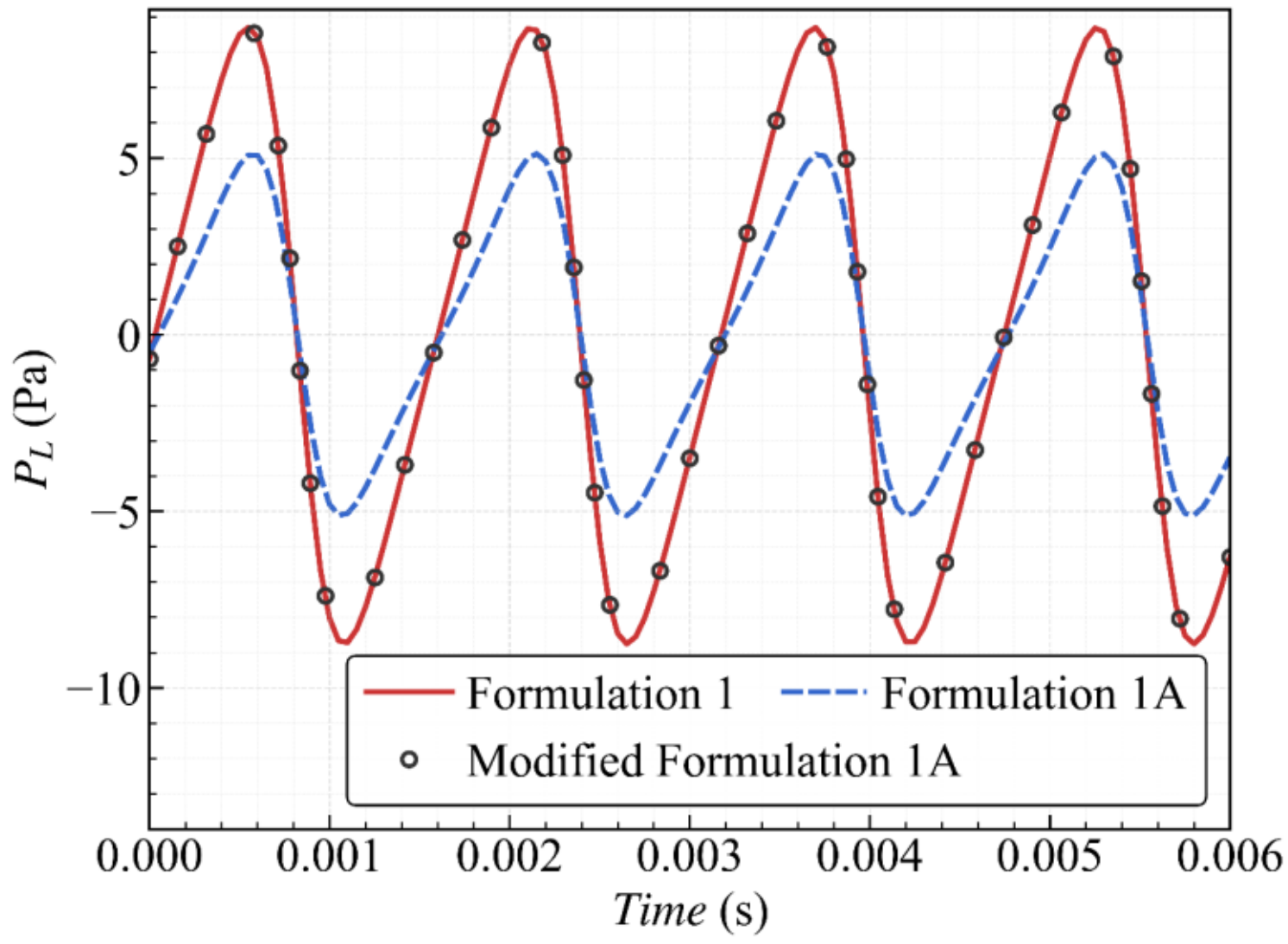

Fig. 3 Predicted time-domain acoustic pressure for multiple formulations

Figure 4 illustrates the contributions of the individual terms in the present

Modified Formulation 1A. Since steady CFD data are employed as input, the TDP term is absent. The magnitude of the TDNV term is comparable to that of the TDMA term, and exceeds that of the TDR term. Although the relative magnitudes and phases of these contributions may vary with source characteristics, the TDNV contribution remains non-negligible.

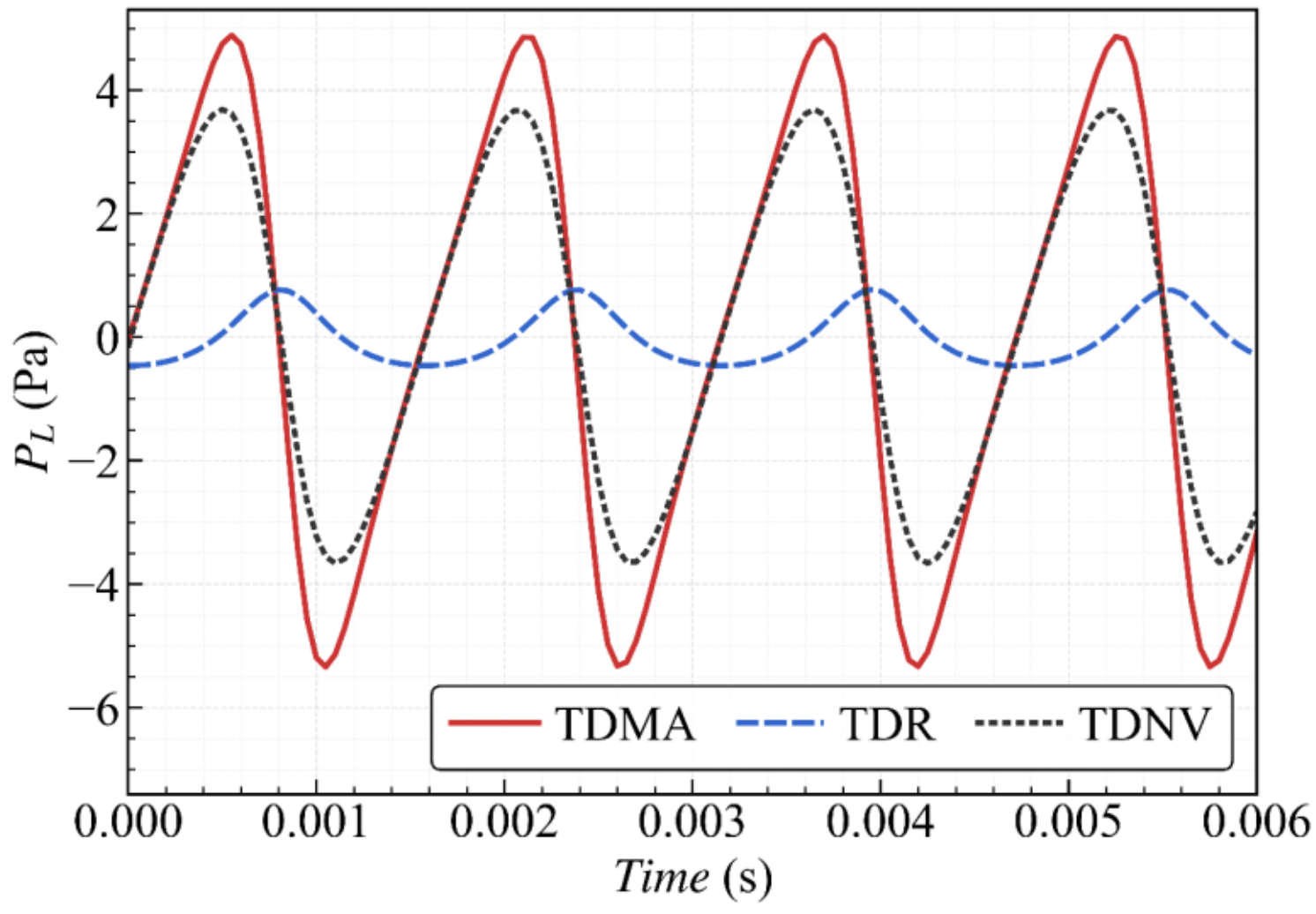

Fig. 3 Contributions of individual terms in the Modified Formulation 1A

By applying a discrete Fourier transform (DFT) to the time-domain signal associated with the TDNV term, its frequency-domain representation is obtained. The sound pressure level (SPL) at the fundamental frequency, i.e., the first blade passing frequency (1BPF), is then extracted to quantify the TDNV contribution. Figure 5 presents the variation of the TDNV contribution to the SPL at 1BPF with rotational speed. Consistent with the theoretical analysis, the TDNV contribution is negligible at low rotational speeds (18.45 dB at 1000 rpm), but becomes increasingly significant as the rotational speed increases (109.61 dB at 10000 rpm). Consequently, for rapidly rotating acoustic sources such as propellers, the TDNV term constitutes an important component of far-field loading noise.

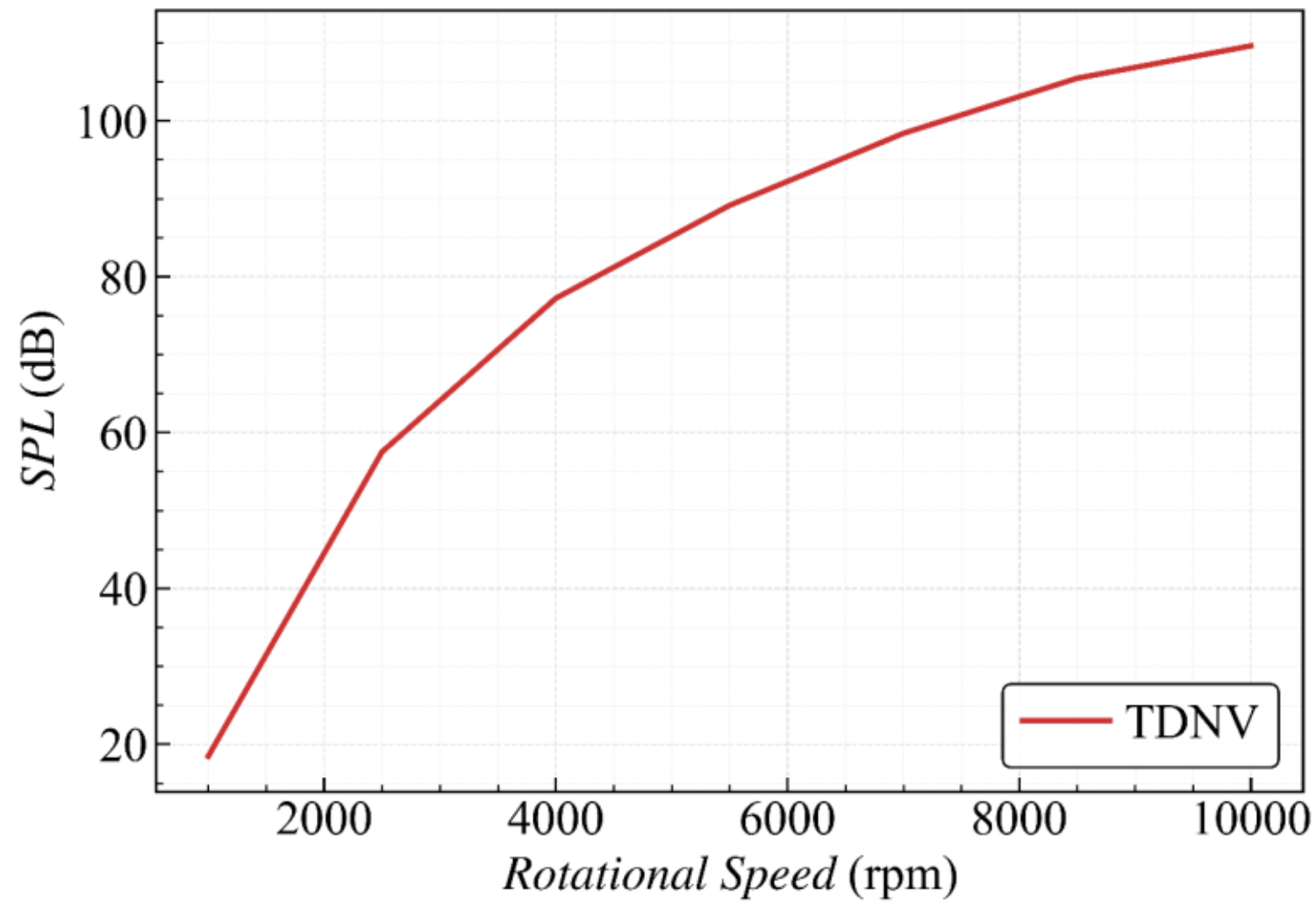

Fig. 4 Variation of the TDNV contribution to SPL at 1BPF with rotational speed

## 4 Conclusions

The derivation of Farassat's Formulation 1A omits the contribution associated with variations in the source surface normal vector. In the present study, the necessity of retaining this contribution is established through a rigorous mathematical derivation. A new time-domain formulation, termed the Modified Formulation 1A, is proposed in which the temporal derivative of the source surface normal vector is explicitly retained. Analysis and numerical validation demonstrate that, except for rigid source surfaces undergoing pure translation, the TDNV contribution must be accounted for in both theoretical analyses and numerical predictions of far-field noise. The Modified Formulation 1A proposed in the present study preserves the computational efficiency of Formulation 1A while extending its applicability to deforming surfaces and rigid bodies undergoing general three-dimensional motion.

## 5 Acknowledgement

This investigation is supported by National Natural Science Foundation of China (Grant Nos. 52306056 and 52476032) and an open research subject of Key Laboratory of Aerodynamic Noise Control (No. ANCL20230201), which are greatly appreciated.